\documentstyle[12pt]{article}

\def\title#1#2#3#4#5{
        \begin{center} \begin{tabular}[t]{l} #1 \end{tabular}
        \hfill	\begin{tabular}[t]{r} #2 \end{tabular}
        \\[1cm] {\LARGE\bf #3} \\[.5in] {#4{}}
        \end{center} \vfill \centerline{{\large ABSTRACT}}
					   {\nopagebreak \noindent\begin{quotation}\noindent {\small #5}
								\end{quotation}} \vfill \newpage
        \def\thefootnote{\sharp\arabic{footnote}}}

\begin {document}

\thispagestyle{empty} \title {Febuary 1994} {\bf OITS-534} {Predictive
fermion mass matrix ansatzes in non-supersymmetric SO(10) grand
unification} { {\bf N.G. Deshpande and E. Keith}\\ {\em Institute of
Theoretical Science\\ University of Oregon, Eugene, OR  97403-5203}}
{We investigate the status of predictive fermion mass ansatzes which
make use of the grand unification scale conditions
$m_e=m_d/3$, $m_\mu =3m_s$, and $\mid V_{cb}\mid =\sqrt{m_{c}/m_{t}}$
in non-supersymmetric SO(10) grand unification. The gauge symmetry
below an intermediate symmetry breaking scale
$M_I$ is assumed to be that of the standard model with either one Higgs
doublet or two Higgs doublets . We find in both cases that a maximum of
5 standard model parameters may be predicted within $1\sigma$
experimental ranges. We find that the standard model scenario predicts
the low energy
$\mid V_{cb}\mid$ to be in a range which includes its experimental
mid-value 0.044 and which for a large top mass can extend to lower
values than the range resulting in the supersymmetric case. In the two
Higgs standard model case, we identify the regions of parameter space
for which unification of the bottom quark and tau lepton Yukawa
couplings is possible at grand unification scale. In fact, we find that
unification of the top, bottom and tau Yukawa couplings is possible
with the running b-quark mass within the
$1\sigma$ preferred range $m_b=4.25\pm 0.1\, GeV$ provided
$\alpha_{3c}(M_Z)$ is near the low end of its allowed range. In this
case, one may make 6 predictions which include $\mid V_{cb}\mid$ within
its $90\%$ confidence limits. However unless the running mass
$m_b>4.4\, GeV$, third generation Yukawa coupling unification requires
the top mass to be greater than
$180\, GeV$. We compare these non-supersymmetric cases to the case of
the minimal supersymmetric standard model embedded in the SO(10) grand
unified group. We also give an example of a possible mechanism, based
on induced vacuum expectation values and a softly broken
$U(1)^3$ symmetry for generating the observed heirarchy of masses and a
mass matrix texture. }

\section{Introduction} Recently, much attention has been given to the
successes of predictive ansatzes \cite{DHR,ADHRS,Barger,otherthanDHR}
for the fermion sector of the standard model (SM). Although originally
fermion sector ansatzes
\cite{Fritsch,GJ,egSM,morerecent,RRRso10} were proposed for and used in
non-supersymmmetric SM \cite{egSM,morerecent}, SU(5)  and SO(10)
\cite{so10} grand unified models, the recent attention has focused on
the case of the minimal supersymmetric standard model (MSSM) contained
in supersymmetric SO(10). One reason for using the ansatze in the
context of a grand unified theory is that in these theories the masses
of the down quarks and the charged leptons are necessarily related.
This gives the possibility of increased predictive ability which, for
example, may be realized in the Georgi-Jarlskog  (GJ) mechanism
\cite{GJ} which has at grand unification scale $m_e = {m_d / 3}$,
$m_\mu = 3 m_s$ and $m_\tau =m_b$. Also, there is the possibility of
relating the up quark mass matrix to the down quark mass matrix
\cite{ADHRSII}.  This happens when the up and down quarks receive their
masses from the same Yukawa couplings or higher dimensional operators
in the context of the grand unified theory. It has also been shown
\cite{DHR,otherthanDHR} that by applying an ansatze with $\mid
V_{cb}\mid = \sqrt{m_c / m_t}$ at grand unification scale, and
requiring the zero terms in the mass matrices to be protected by some
symmetries above grand unification scale, $\mid V_{cb}\mid$ is
predicted to be within or close to the upper end of the
$1\sigma$ experimental range with out requiring $m_t$ to be too large.
SO(10) (or a group like $E_6$ containing SO(10)) is the chosen group
because then, unlike with SU(5), the mass matrices can be automatically
symmetric, neutrinos may be given small masses with mixing to solve the
solar neutrino problem, and there are useful relations between the mass
matrices
\cite{DHR}. In the DHR (Dimopolous-Hall-Raby) formulation \cite{DHR},
the MSSM with gauge coupling unification is chosen because by requiring
unification of gauge couplings and the supersymmetry (SUSY) effective
scale parameter $M_S$ to be in the proximity of $1\, TeV$, as is needed
for SUSY to solve the fine-tuning problem, one can predict
$\alpha_{3c}(M_Z)$ to be within its experimentally determined range
from the experimentally well determined parameters $\alpha$ and
$\sin{\theta_W}$ \cite{gaugeprediction}.

Although the fermion mass ansatzes in SUSY SO(10) have so far worked
quite well, there is, as of yet, no evidence for SUSY and one may wish
to compare the predictions and predictive ability of ansatzes with SUSY
to those without SUSY. This is useful not only because we do not know
whether SUSY exists, but also because many parameters of the fermion
mass and the quark mixing sector have not yet been determined with
great precison, so we can not yet be confident of the success of the
predictions of any particular scheme. The first comprehensive
discussion of the predictions in the fermion sector of an ansatze was
done in ref \cite{DHR} for the case of MSSM contained in SUSY SO(10).
Only recently, has the low energy data (LED) been precise enough to
give a reasonable test of the predictions of an ansatze. In this paper,
we will look at fermion mass ansatzes in non-SUSY SO(10) grand
unification in terms of current LED.

As in the paper of ref. \cite{DHR}, we take the ansatze at unification
scale and assume that some, as yet, unspecified symmetries enforce the
zero terms in the fermion mass matrices at that scale. One expects that
such symmetries originate in a theory that is realized at scales equal
to or greater than the grand unification scale and that these
symmetries are broken at the grand unification scale, which allows the
zero terms in the fermion mass matrices to develop finite values from
renormalization group effects. We will suggest an example of such a
scenario in Section 6 of this paper. Without the intention of examining
all possible textures of fermion mass matrices, we will assume an up
quark mass matrix based on the Fritsch ansatze
\cite{Fritsch} and down and charged lepton mass matrices based on the
Georgi-Jarlskog ansatze \cite{GJ}. Ansatzes of this general form have
been used extensively in the literature.

Although SUSY SO(10) can break to the MSSM in only one step, non-SUSY
SO(10), in general, needs at least two steps to break to SM. Typically,
in two step breaking of SO(10) to SM with Higgs particles taking masses
according to the principal of minimal fine-tuning \cite{finetuning}, the
intermediate scale $M_I\sim 10^{9}$ to
$10^{11}\, GeV$ and the unification scale
$M_U\sim 10^{16}\, GeV$ \cite{so10results}. The allowed single
intermediate scale gauge symmetries are the four groups $2_L\, 2_R\,
4_C$, $2_L\, 2_R\, 4_C\, P$, $2_L\, 2_R\, 1_{B-L}\, 3_c$  and $2_L\,
2_R\, 1_{B-L}\, 3_c\, P$, where $P$ refers to D-parity not having been
broken. (Only in SUSY SO(10) is
$SU(5)
\times U(1)$ as an intermediate symmetry group possible.) Another
possibility, pointed out recently, is that if threshold effects are not
minimized \cite{MOHAthreshold}, but to the contrary super heavy Higgs
particles not contributing to proton decay are allowed to vary below a
SM coupling unification scale by a factor that can be as high as 10,
then it is possible for ${M_U/ M_I}\leq 30$ \cite{wolf}. Like the SUSY
case, this scheme makes one low energy prediction in the gauge sector
from two inputs. It predicts
$\alpha_{3c} (M_Z)$ in the range of 0.119 to 0.125.  In our paper, we
will look at cases where SO(10) breaks at a scale $M_U$ via the VEV
contained a
$\bf{210}$ \cite{twoten} representation Higgs to the gauge  symmetry
$2_L\, 2_R\, 4_C$ and next at a scale $M_I\sim 10^{11}$ or $10^{14}\,
GeV$ to the SM.  Further, we will assume that the vacuum expectation
value (VEV) which breaks the gauge symmetry $2_L\, 2_R\, 4_C$ to the SM
is contained in an
$SU(2)_R$ triplet of a $\bf{126}$ representation Higgs field. This
gives the right-handed neutrinos Majorana masses. As is usual, we use
the VEV of a complex ${\bf 10}$ representation Higgs field for the
electroweak symmetry breaking. Even though the scheme of ref.
\cite{wolf} requires high values of
$\alpha_{3c} (M_Z)$, we will consider $\alpha_{3c} (M_Z)=0.118\pm
0.007$ for both
$M_I\sim 10^{11}$ and $M_I\sim 10^{14}\, GeV$.

Below the scale $M_I$, we consider two possibilities, one that the
effective theory is the conventional one Higgs doublet SM and the
second possibility that the effective theory is the two Higgs standard
model (2HSM). The reason we are interested in the 2HSM is that, while
as we will see in the SM that the   unification of the Yukawa couplings
of the bottom quark and tau lepton is not feasible, both the
unification of the Yukawa couplings of the bottom quark and tau lepton
and unification of all three third generation SM Yukawa couplings is
possible in the 2HSM.

The rest of this paper is organized in the following manner. In the next
section, we will discuss the renormalization group equations (RGE's) of
the fermion sector parameters and the gauge couplings. After that, we
review the basic results of implimenting the GJ ansatze in the MSSM.
We do this so that we may later compare the results for the two cases
without SUSY to the case with SUSY. In the fourth section, we will
discuss the case of fermion mass ansatzes when between the scales of
$m_t$ and $M_I$ the effective theory is the SM. In the fifth section,
we discuss the case of fermion mass ansatzes when instead of the SM the
effective theory below $M_I$ is  the 2HSM.  Next, we give an example of
a possible explanation of fermion generation mass heirarchy and flavor
symmetries by use of induced VEV's
\cite{BM} in super heavy Higgs fields and a softly boken $U(1)^3$
symmetry. In the final section, we summarize the paper.

\section{RGE's and LED}

Here, we remind the reader of how Yukawa couplings evolve in the SM
gauge symmetry $1_Y\, 2_L\, 3_c$ in the 1-loop approximation
\cite{YukawaRGE}, which we will use. Let ${\bf U}$, ${\bf D}$, and
${\bf E}$ be the
$3\times 3$ Yukawa matrices in generation space for the up and down
quarks, and the charged leptons, respectively. In the SM, we have the
Yukawa couplings \begin{eqnarray} {\cal L}_Y=\overline{q}_L {\bf U}
\tilde{\phi} u_R\, +\, \overline{q}_L {\bf D} \phi d_R\, +\,
\overline{l}_L {\bf E} \phi e_R\, +\, h.c.\, . \end{eqnarray} In the
MSSM and in the 2HSM we have  \begin{eqnarray} {\cal
L}_Y=\overline{q}_L {\bf U}
\phi_u u_R\, +\, \overline{q}_L {\bf D} \phi_d d_R\, +\, \overline{l}_L
{\bf E} \phi_d e_R\, +\, h.c.\, ,
\end{eqnarray} where $\left< |\phi_u |\right> =\kappa_u$ and $\left<
|\phi_d |\right> =\kappa_d$ with $\sqrt{\mid \kappa_u \mid^2 + \mid
\kappa_d
\mid^2} =\kappa=174\, GeV$ and ${\kappa_u/ \kappa_d}\equiv
\tan{\beta}$. The 1-loop RGE's for these couplings are \begin{eqnarray}
16\pi^2 {d {\bf U}\over dt}=[ Tr( 3{\bf U} {\bf U}^\dagger \, +\,
3a{\bf D} {\bf D}^\dagger
\, +\, a{\bf E E}^\dagger )
\nonumber \\ +\, {3 \over 2} ( b{\bf U U}^\dagger \, +\, c{\bf D
D}^\dagger )
\, -\, \Sigma c_i^{(u)} g_i^2 ] {\bf U}\, ,\\ 16\pi^2 {d {\bf D}\over
dt}=[ Tr ( 3a{\bf U U}^\dagger \, +\, 3{\bf D D}^\dagger \, +\, {\bf E
E}^\dagger )
\nonumber \\ +\, {3\over 2} ( b{\bf D D}^\dagger \, +\, c{\bf U
U}^\dagger )
\, -\, \Sigma c_i^{(d)} g_i^2 ] {\bf D}\, ,\\ 16\pi^2 {d {\bf E}\over
dt}=[ Tr ( 3a{\bf U U}^\dagger \, +\, 3{\bf D D}^\dagger \, +\, {\bf E
E}^\dagger )
\nonumber \\ +\, {3\over 2}b{\bf E E}^\dagger \, -\, \Sigma c_i^{(e)}
g_i^2 ] {\bf E}\, ,	\end{eqnarray} with $t=\ln \mu$, \begin{eqnarray}
\rm{SM}:\, \, \, (a,b,c) = (1,1,-1)\, ,\\ \rm{2HSM}:\, \, \, (a,b,c) =
(0,1,{1\over 3})\, ,\\ \rm{MSSM}:\, \, \, (a,b,c) = (0,2,{2\over 3})\, ,
\end{eqnarray} and \begin{eqnarray}
\rm{SM;2HSM}:\, \, \, c_i^{(u)} = ({17 \over 20},{9 \over 4},8)\, ,\,
c_i^{(d)}=({1 \over 4},{9 \over 4},8) \, , \, c_i^{(e)}=({9\over
4},{9\over 4},0)\, ,\\ \rm{MSSM}:\, \, \, c_i^{(u)} = ({13 \over
15},3,{16\over 3})\, ,\, c_i^{(d)}= ({7
\over 15},3,{16 \over3}) \, ,\, c_i^{(e)}=({9\over 5},3,0)\, .
\end{eqnarray}

In computing the evolution of the gauge couplings, we will use a 2-loop
analysis  but we will ignore the small effects of the Yukawa couplings
on their running. The two loop equations, which we numerically
integrate, are of the form \begin{eqnarray}
\mu{\partial \alpha_i^{-1} (\mu)\over \partial \mu}=-{1\over 2\pi}
\left( b_i+{b_{ij}\over 4\pi}\alpha_j^{} (\mu) \right) \, .
\end{eqnarray} The 1-loop coefficients $b_i$ are \begin{eqnarray}
\rm{SM}:\, \, \, (b_1,b_2,b_3) = ({41\over 10},{-19\over 6},-7)\, ,\\
\rm{2HSM}:\, \, \, (b_1,b_2,b_3) = ({21\over 5},3,-7)\, ,\\
\rm{MSSM}:\, \, \, (b_1,b_2,b_3) = ({33\over 5},1,-3)\, .
\end{eqnarray} The two loop coefficients $b_{ij}$ can be extracted from
ref.
\cite{twoloopgauge}. We use gauge couplings normalized so as to become
equal at the scale
$M_U$. We use the following gauge sector inputs
\cite{gaugeinput}:
\begin{eqnarray} \alpha^{-1} (M_Z) &=& 127.9\, ,\nonumber\\
\alpha_{3c} (M_Z) &=& 0.118\pm 0.007\, ,\nonumber\\ \tilde{x} (M_Z) &=&
0.2326\, ,\nonumber\\ M_Z &=& 91.187\pm 0.007\, GeV\, ,
\end{eqnarray} with \begin{eqnarray} \alpha_{1Y}^{-1}(M_Z) &=& {3\over
5}{1-\tilde{x} \over \alpha(M_Z)}\, ,\\
\alpha_{2L}^{-1}(M_Z) &=& {\tilde{x} \over \alpha (M_Z)}\, ,
\end{eqnarray} and we have used the experimental mid-values for
$\alpha(M_Z)$ and $\tilde{x} \equiv \sin^2\theta_W (\overline{MS} )$.

As in ref. \cite{DHR},  we numerically integrate $\alpha_1$, $\alpha_2$
and
$\alpha_3$ from $M_Z$ up to a scale $\mu_t$ which is in the vicinity of
where we expect to find the running mass $m_t$ in the
$\overline{MS}$ scheme. Between $\mu=M_Z$ and $\mu=\mu_t$, we use the
2-loop SM gauge evolution with 1-loop threshold corrections for
$m_t=\mu_t$ to find
$\alpha_{1Y}$, $\alpha_{2L}$, and
$\alpha_{3c}$ at the scale $\mu=\mu_t$. From $\mu=\mu_t$ down to a
particular fermion's running mass for $m_b$, $m_c$ or charged leptons or
down to $1\, GeV$ for the less massive quarks, we calculate the running
of its mass according to 3-loop QCD \cite{threeloop} and 1-loop QED
effects. CKM parameters are evaluated at the scale $\mu_t$. Of course,
we always use the effective theory where all fermions more massive than
the scale of interest have been integrated out. These effects are
represented by
$m_i=m_i(\mu_t) \eta_i$. In this report, we take $\mu_t=180\, GeV$, and
find
$\alpha_1^{-1}(\mu_t)=58.51$,
$\alpha_2^{-1}(\mu_t)=30.15$, and $\alpha_3^{-1}(\mu_t)=9.30\pm 0.5$ and
\begin{eqnarray} \eta_b &=& 1.53 ^{+0.07}_{-0.06}\, ,\nonumber\\ \eta_c
&=& 2.20^{+0.27}_{-0.20}\, ,\nonumber\\
\eta_s &=& \eta_d=2.45^{+0.37}_{-0.26}\, ,\nonumber\\ \eta_u &=&
2.46^{+0.37}_{-0.26} \, ,\nonumber\\ \eta_{e} &\approx &
\eta_{\mu} \approx \eta_{\tau}=1.015 \, . \end{eqnarray}

We are interested in the low energy fermion masses, the CKM quark mass
mixing matrix elements $V_{\alpha \beta}$ \cite{LEDRGE}, and the
Jarlskog CP violation parameter
$J$.   In the approximation that we use the 1-loop Yukawa RGE's, ignore
terms
$O(\lambda_c^2)$ or smaller where
$\lambda_i$ is the Yukawa coupling of fermion $i$, and set $\eta_t=1$,
the exact solutions for the LED in terms of the same parmeters at an
intermediate breaking $\mu =\mu_I$ are the following: \begin{eqnarray}
m_t(m_t) &=& m_t(\mu_I) A_u e^{-(3+{3\over 2}b)I_t-(3a+{3\over
2}c)I_b-aI_\tau}\, ,\label{mt}\\ m_b(m_b) &=& m_b(\mu_I) \eta_b A_d
e^{-(3a+{3\over2}c)I_t-(3+{3\over 2}b)I_b-I_\tau} \, ,\\
m_{\tau}(m_{\tau}) &=& m_{\tau}(\mu_I) \eta_{\tau} A_d
e^{-3aI_t-3I_b-(1+{3\over 2}b)I_\tau}\, ,\\ m_c(m_c) &=& m_c(\mu_I)
\eta_c A_u e^{-3I_t-3aI_b-aI_\tau}\, ,\label{mc}\\ m_i(m_i) &=&
m_i(\mu_I) \eta_{\mu} A_e e^{-3aI_t-3I_b-I_\tau}
\,
\, \, \, \, \, \, {\bf{(i=\mu,e)}}\, ,\\ m_i(1\, GeV) &=& m_i(\mu_I)
\eta_s A_d e^{-3aI_t-3I_b-I_\tau} \, \, \, \, \, \, \, {\bf{(i=s,d)}}\,
,\\ m_u(1\, GeV) &=& m_i(\mu_I) \eta_u A_u e^{-3I_t-3aI_b-aI_\tau}\,
,\\	\mid V_{\alpha
\beta}(m_t)\mid &=&\mid V_{\alpha \beta}(\mu_I)\mid e^{{3\over
2}cI_t+{3\over 2}cI_b} \, \, \, \, \, \, \, {\bf{(\alpha \beta =
ub,cb,tb,ts)}}\, ,\label{vcb}\\ \mid V_{\alpha \beta}(m_t)\mid &=& \mid
V_{\alpha
\beta}(\mu_I)\mid \, \, \, \, \, \, \, {\bf{(other\, \alpha \beta)}}\,
,\\ J(m_t) &=& J(\mu_I )e^{3cI_t+3cI_b}\, , \end{eqnarray} where the
effect of third generation Yukawa couplings on the Yukawa evolution is
given as
\cite{ADHRSII}:
\begin{eqnarray}  I_i= \int_{\mu_t}^{\mu_I}\left( \lambda_i
\over 4\pi \right)^2 dt\, , \end{eqnarray}  and the effect of gauge
couplings on Yukawa evolution is given as \cite{JohnH} \begin{eqnarray}
A_\alpha=\exp \left[ {1\over 16\pi^2}\int_{\ln\mu_t}^{\ln\mu_I}\Sigma
c_i^{(\alpha )}g_i^2(\mu )d(\ln\mu)\right]\, .\label{gaugecont}
\end{eqnarray}  In the 1-loop approximation for the gauge RGE's
$A_\alpha$ becomes
\begin{eqnarray} A_\alpha=\prod \left( {\alpha_{i_I}\over
\alpha_{i_t}}\right)^{c_i^{(\alpha )}\over 2b_i}\, .
\end{eqnarray} In the SM or in the 2HSM or MSSM when
$\tan{\beta}$ is small, it is a very good approximation to ignore terms
$O(\lambda_b^2)$ in the Yukawa coupling evolution equations, in which
case
\cite{JohnH} \begin{eqnarray} e^{I_t}=\left[ 1+\lambda_{t}(M_I)^2
K_u^{(tI)}
\right]^{1\over 6+3b}\label{Ku}
\, ,\end{eqnarray}where \begin{eqnarray} K_u={6+3b\over
16\pi^2}\int_{\ln\mu_t}^{\ln\mu_I} \exp \left[ {1\over
8\pi^2}\int_{\ln\mu '}^{\ln\mu_I}\Sigma c_i^{(\alpha )}g_i^2(\mu
')d(\ln\mu ')\right] d(\ln
\mu)\, .\label{K} \end{eqnarray}

In Table 1, we give the values for the $A_\alpha$'s and the $K_u$'s for
the SM and the 2HSM. We show two different cases for the situation
where the effective theory below the scale $M_I$ is the SM. In the SM
case (a)
$M_I=10^{10.94}\, GeV$, and in the SM case (b) $M_I=10^{14}\, GeV$. In
the case where the effective theory between $\mu_t$ and $M_I$ is the
2HSM, we use $M_I=10^{11.28}\, GeV$.  Note that the $A_\alpha$'s and
the $K_u$'s in the SM case (a) and the 2HSM case have very similar
values.  For the sake of comparison, we also show the $A_\alpha$'s and
$K_u$ for the case when the effective theory above the scale
$\mu_t=180\, GeV$ is the MSSM. In this case, the upper bound of
integration in the $A_\alpha$'s and $K_u$ is the gauge coupling
unification scale $M_U$. The strong coupling constant
$\alpha_{3c}(M_Z)=0.121$ is determined by requiring gauge coupling
unification to be acheived with $\alpha$ and $\sin{\theta_W}$ as
inputs.

In Table 1, we also show the ratio $A_d/A_e$ in the different cases
because the ratio of the masses of the down quarks to the masses of the
charged leptons is proportional to $A_d/A_e$. Note that this ratio is
highest in the MSSM scenario. In the SM case (b) this ratio is higher
than in the other two non-SUSY cases because the $SU(4)_C$ gauge
symmetry is broken at $M_I$, which  for SM case (b) is larger than for
the other two non-SUSY scenarios considered.

We can use the $A_u$'s and $K_u$'s of Table 1 to find the infrared
quasi-fixed point of the top quark \cite{fixedpoint}. When
$\lambda_t>>\lambda_b$,
\begin{eqnarray}
\lambda_t={A A_u\over \sqrt{1+A^2 K_u}}\, ,
\end{eqnarray} where $A$ is the top quark Yukawa coupling at the scale
$M_I$ for the non-SUSY cases and at $M_U$ for the MSSM case. In the
limit of a large $A$, one finds $\lambda_t\approx A_u/\sqrt{K_u}$.
Therefore in the MSSM when $\sin{\beta}\approx 1$ and
$\lambda_t>>\lambda_b$, $ \left( A_u/\sqrt{K_u}\right) \kappa$ is the
infrared quasi-fixed point of the top quark. For the MSSM case, one
finds that the fixed point is $194\, GeV$. This gives an upper bound
for the running mass $m_t$ for any $\tan{\beta}$.

However when an intermediate breaking scale $M_I$ exists, $A$ has an
upper bound from the following equation which is valid when the
intermediate gauge symmetry is
$2_L\, 2_R\, 4_C$:
\begin{eqnarray} A={\lambda_{t_U} A_f\over \sqrt{1+\lambda_{t_U}^2
K_f}}\, ,
\label{lambdatI}
\end{eqnarray} where we have defined the effect of the intermediate
scale gauge couplings $g_{2L}$, $g_{2R}$, and $g_{4C}$ on the Yukawa
coupling evolution of all fermions as
\begin{eqnarray} A_f=\exp \left[ {1\over
16\pi^2}\int_{\ln\mu_t}^{\ln\mu_I}\Sigma c_i^{(f)}g_i^2(\mu
)d(\ln\mu)\right]\, ,
\end{eqnarray} and defined the analog of $K_u$ as
\begin{eqnarray} K_f={3\over 4\pi^2}\int_{\ln\mu_t}^{\ln\mu_I} \exp
\left[ {1\over 8\pi^2}\int_{\ln\mu '}^{\ln\mu_I}\Sigma
c_i^{(f)}g_i^2(\mu ')d(\ln\mu ')\right] d(\ln \mu)\, ,
\end{eqnarray} with
\begin{eqnarray} c_i^{(f)}=\left( {9\over 4},{9\over 4},{45\over
4}\right)
\, ,
\end{eqnarray} and $\lambda_{t_U}$ is the top quark Yukawa coupling at
$M_U$.  Eq\@. (\ref{lambdatI}) is the solution to the intermediate scale
equation
\begin{eqnarray} 16\pi^2{d\ln{\lambda_t}\over dt}=\left(
6\lambda_t^2-\Sigma c_i^{(f)}g_i^2
\right) \, .
\end{eqnarray} For the SM case (a), we find the fixed point to be
$223\pm 3\, GeV$. For the SM case (b), we find $\kappa
A_u/\sqrt{K_u}=235\pm 4\, GeV$. For the 2HSM case, we find the upper
bound of the top running mass to be $225\pm 3\, GeV$. As is well known,
without SUSY the fixed point of the top quark is clearly higher than
that allowed for by examination of electroweak data \cite{topmass}.

We now should consider the relations between $m_b$ and  $m_\tau$ in the
three cases. They are
\begin{eqnarray} {m_b\over m_\tau}&=&{\lambda_{b_U}\over
\lambda_{\tau_U}}{\eta_b\over \eta_\tau} {A_d \over A_e} e^{-{3\over
2}cI_t-{3\over 2}bI_b+{3\over 2}bI_\tau}\, ,\\  &=&{\lambda_{b_U}\over
\lambda_{\tau_U}}{\eta_b\over \eta_\tau} {A_d\over A_e} e^{{3\over
2}I_t-{3\over 2}I_b+{3\over 2}I_\tau} \,
\, \, \, (SM)\, ,\\ &=&{\lambda_{b_U}\over \lambda_{\tau_U}}{\eta_b\over
\eta_\tau} {A_d\over A_e} e^{-{1\over 2}I_t-{3\over 2}I_b+{3\over
2}I_\tau}
\, \, \, \, (2HSM)\, , \label{mbtwo}\\ &=&{\lambda_{b_U}\over
\lambda_{\tau_U}}{\eta_b\over \eta_\tau} {A_d\over A_e}
e^{-I_t-{3}I_b+{3}I_\tau} \, \, \, \, (MSSM)\label{mbmssm}\, ,
\end{eqnarray}where the subscript $U$ on a parameter denotes its value
at unification scale. The $SU(3)_c$ gauge contribution by itself would
make
$m_b$ undesirably large for the case of bottom-tau Yukawa coupling
unification with the requirement $m_\tau =1.784\, GeV$.

In the SM case, the ratio $m_b/m_\tau$ increases with top quark mass.
For example, if we assume $m_b=m_\tau$ at grand unification scale and
use
$m_\tau=1.784\, GeV$ as an input, then the lowest possible value of
$m_b$ is obtained for the lowest reasonable values of $m_t$,
$\alpha_{3c}(M_Z)$, and $M_I$, which are pole mass $m_t\approx 130\,
GeV$,
$\alpha_{3c}(M_Z)=0.111$, and $M_I\sim 10^{11}\, GeV$. This gives a
running mass $m_b= 5.0\, GeV$ or $m_b^{pole}= 5.2\, GeV$. This $m_b$ is
too large to be acceptable . Because of this, we are forced into using
two Yukawa couplings to give mass to the bottom and tau fermions in the
one Higgs case. One coupling must be to a $\bf{10}$ representation
Higgs and the other to a
$\bf{\overline{126}}$ representation Higgs. (Remember that, unlike a
coupling to a $\bf{10}$, couplings to
$\bf{\overline{126}}$'s contribute to lepton Dirac masses relative to
quark masses with a factor of the Clebsch $-3$.) We assume the entire
bidoublet of the $\bf{ \overline{126}}$ representation Higgs field to
have a mass of the order of $M_U$ and to contribute to the fermion
masses through a VEV induced from the VEV of the
${\bf 10}$ representation Higgs field
\cite{BM}.

On the other hand, in the 2HSM and the MSSM when we input
$m_\tau=1.784$ and require the unification scale condition
$m_b(M_U)=m_\tau(M_U)$, the ratio $m_b/m_\tau$ decreases with increasing
$m_t$. Bottom-tau Yukawa coupling unification has proved successful in
the MSSM. We will see later that this is also possible in the 2HSM,
although the fit is not as attractive. This is because the ability of
the top quark Yukawa coupling to keep the ratio
$m_b/ m_\tau$ from becoming too large is less in the 2HSM than in the
MSSM.

Since we are interested in matrices of the GJ form which have
$\mid V_{cb}(M_U)\mid =\sqrt{m_{c}(M_U)/ m_{t}(M_U)}$, we  also
consider the equations \begin{eqnarray} {\mid V_{cb}
\mid^2\over \left( {{m_c\over m_t}}\right)}&=&\eta_c^{-1} e^{(-{3\over
2}b+3c)I_t+{3\over 2}cI_b}\, \\ &=&\eta_c^{-1} e^{-{9\over 2}I_t-{3\over
2}I_b}\, \, \, \, (SM)\, ,\\ &=&\eta_c^{-1} e^{-{1\over 2}I_t+{1\over
2}I_b}\, \, \, \, (2HSM)\, ,\label{vcbtwo}\\ &=&\eta_c^{-1}
e^{-I_t+I_b}\,
\, \, \, (MSSM)\label{vcbmssm}\, . \end{eqnarray} We see that in all
cases, the heavier the top quark is, the lower this ratio is.

\section{Brief Review of MSSM case (DHR Ansatze)} In this section, we
will look at the ansatze of Dimopolous, Hall, and Raby (DHR)
\cite{DHR} for the purpose of making the program we will use for the
non-SUSY cases clear and also so that we may later compare results
between the SUSY and non-SUSY cases. For a more complete analysis, see
ref.
\cite{DHR,ADHRS,Barger}. In the original DHR ansatze, the the grand
unification scale fermion Yukawa coupling matrices take the following
form:   \begin{eqnarray} {\bf U}= \left( \matrix{ {0}&{C}&{0}\cr
{C}&{0}&{B}\cr {0}&{B}&{A}\cr }\right)\, ,
\, {\bf D}= \left(
\matrix{ {0}&{F}&{0}\cr {F}&{E}&{0}\cr {0}&{0}&{D}\cr }\right)\, ,\,
{\bf E}=
\left( \matrix{ {0}&{F}&{0}\cr {F}&{-3E}&{0}\cr {0}&{0}&{D}\cr
}\right)\, ,
\label{GJ} \end{eqnarray} where $A$, $B$,$C$, $D$, $E$, and $F$ are
complex parameters, with
\begin{eqnarray}|A|>>|B|>>|C|\nonumber\\|D|>>|E|>>|F|\,
.\label{hierarch}\end{eqnarray} (Note that the up-quark mass matrix is
of the Fritzsch form and that the down-quark and charged-lepton mass
matrices impliment the Georgi-Jarlskog mechanism.) We recall that
$M_U=U \kappa
\sin{\beta}$, $M_D=D \kappa \cos{\beta}$, and $M_E=E \kappa
\cos{\beta}$. After rotating away all but one unavoidable phase $\phi$
in the Yukawa coupling  matrices by redefinition of the phases of the
fermion fields
\cite{DHR}, these matrices may be given the following form:
\begin{eqnarray} {\bf U}= \left( \matrix{ {0}&{C}&{0}\cr {C}&{0}&{B}\cr
{0}&{B}&{A}\cr }\right)\, ,\, {\bf D}= \left( \matrix{
{0}&{Fe^{i\phi}}&{0}\cr {Fe^{-i\phi}}&{E}&{0}\cr {0}&{0}&{D}\cr
}\right)\, ,\, {\bf E}= \left( \matrix{ {0}&{F}&{0}\cr {F}&{-3E}&{0}\cr
{0}&{0}&{D}\cr }\right)\, ,\label{rotatedaway} \end{eqnarray} where
$A$, $B$,$C$, $D$, $E$, and $F$ are now real.  This ansatze uses the 8
inputs $A$, $B$, $C$, $D$,
$E$, $F$, $\phi$, and
$\tan{\beta}$ to describe the SM fermion sector, which contains 13
independent parameters. Hence, these 8 parameters may be fixed in terms
of the 8 best measured SM fermion sector parameters to yield 5 SM
fermion sector predictions and  $\tan{\beta}$ of the MSSM. The
following inputs are used \cite{experiment}:
\begin{eqnarray} m_b(m_b)&=&4.25\pm0.1\, GeV\, ,\\
m_{\tau}(m_{\tau})&=&1.784\, GeV\, ,\\ m_c(m_c) &=& 1.27\pm 0.05\,
GeV\, ,\\ m_\mu (m_\mu ) &=&105.658\, MeV\, ,\\ 0.2 &\leq & {m_u(1\,
GeV)\over m_d(1\, GeV)} \leq 0.7\, ,\label{uoverd}\\ m_e(m_e) &=&
.511\, MeV\, ,\\
\mid V_{cb}\mid &=& 0.044\pm 0.014\, ,\\
\mid V_{us} \mid &=& 0.221\pm 0.003\, .
\end{eqnarray} The above masses are running masses in the
$\overline{MS}$ scheme and their quoted uncertainties are at the
$1\sigma$ level. For the CKM matrix parameters $| V_{cb}|$ and $|
V_{us} |$, we have quoted the uncertainties at the $90\%$  confidence
level. The $1\sigma$ limit on
$| V_{cb}|$ is $| V_{cb}| =0.044\pm 0.009$.

By finding the biunitary transformations that transform the mass
matrices at grand unification scale to diagonal matrices with real
positive entries, making use of Eq\@. (\ref{hierarch}), and using the
results of the RGE analysis of the previous section one may find the
predictions \cite{DHR} for the 5 SM parameters and $\tan{\beta}$ in
terms of the previously given inputs. Four of these are the following:
\begin{eqnarray} {m_d/m_s\over
\left[ 1-{m_d\over m_s } \right]^2}&=& {9{m_e\over m_\mu}\over \left[
1-{m_d\over m_s } \right]^2}\, ,\label{i}\\ m_s-m_d&=&{m_\mu \over
3}{\eta_s\over \eta_\mu}R_{d\over e}\, ,\label{ii}\\ \left| {V_{ub}\over
V_{cb}}\right| &=&\sqrt{3m_e\left( {m_u\over m_d} \right) \over m_c}
{\eta_s
\eta_c\over \eta_u \eta_e}R_{d\over e}\, ,\label{iii}\\ J
&=&\sqrt{m_d\over m_s} | V_{cb}|^2 \left| {V_{ub}\over V_{cb}}\right|
\sin{\phi}\label{iv}\, ,
\end{eqnarray} with
\begin{eqnarray}
\cos{\phi}={| V_{us}|^2 - \left( {m_d\over m_s} \right)-
\left| {V_{ub}\over V_{cb}}\right|^2\over 2\sqrt{m_d\over m_s} \left|
{V_{ub}\over V_{cb}}\right| }\, ,
\end{eqnarray} and where we have defined $R_{d\over e}\equiv A_d/A_e$.

The fifth predicted SM parameter is $m_t$. An input value for $\mid
V_{cb}\mid$ gives two possible pairs of predictions for $m_t$ and the
MSSM parameter $\tan{\beta}$. Only for the case that $\tan{\beta}$ is
small can an accurate analytical approximation be given for $m_t$ and
$\tan{\beta}$. Otherwise, one must numerically integrate the RGE's. When
$\tan{\beta}$ is assumed to be small, the following predictions can be
made from the $M_U$ scale conditions $|V_{cb}|=\sqrt{m_c/m_t}$ and
$m_b=m_\tau$ with the RGE's given in the last section:
\begin{eqnarray} m_t(m_t)&=&{m_c/\eta_c \over \mid V_{cb}\mid^2}{b\over
\tau}\,  ,\\
\sin{\beta}&=&{\sqrt{K_u}\over A_u \kappa}{m_c/\eta_c \over \mid
V_{cb}\mid^2}
\left( {\tau \over b}\right)^{5} \left[ \left( {\tau \over
b}\right)^{12}-1\right]^{-{1\over 2}}\, ,
\end{eqnarray} and for the unification scale top quark Yukawa coupling
\begin{eqnarray} A=K_u^{-{1\over 2}}\sqrt{\left( {\tau \over
b}\right)^{12}-1}
\end{eqnarray}    where we have defined
\begin{eqnarray}
\tau &=&{m_\tau \over \eta_\tau A_e}\, ,\\ b &=&{m_b \over \eta_b
A_d}\, ,
\end{eqnarray} and $m_t$ is the running mass. As is well known, the
$\overline{MS}$ scheme running mass is  related to the physical pole
mass by the relation
\begin{eqnarray} m_t^{pole}=m_t\left( 1+{4\alpha_{3}(m_t)\over 3\pi
}+O(\alpha_{3}^2(m_t)) \right)\, .
\end{eqnarray}

Now, we need to know what ranges of values are acceptable for the output
parameters. For the purpose of comparing later with the non-SUSY cases,
we will give the results for the previously mentioned example of
$M_S=180\, GeV$ and
$\alpha_{3c}(M_Z)=0.121$.  For this value of $\alpha_3(M_Z)$, we find
$\alpha_3(\mu_t)=0.110$ and  the following $\eta_i$'s: \begin{eqnarray}
\eta_b = 1.56\, ,\\
\eta_c= 2.30\, ,\\
\eta_s= 2.58\, ,\\
\eta_u= 2.60\, .\\
\end{eqnarray}

For the outputs $m_s/m_d$ and $m_s$, acceptable ranges are the following
\cite{experiment}: \begin{eqnarray} 15&\leq &{m_s(1\, GeV)\over m_d(1\,
GeV)}\leq25\, ,\label{soverd}\\ m_s(1\, GeV)&=&175\pm 55\, MeV\, .
\end{eqnarray} In ref. \cite{experiment}, larger values of $m_s/m_d$
correspond to smaller values of $m_u/m_d$. Determined solely by the
ratio
$m_e/m_\mu$, the prediction for $m_s/m_d$ is
\begin{eqnarray}  {m_s\over m_d}=24.71\, ,
\end{eqnarray}  which is at the upper end of its acceptable range. (Of
course, this ratio does not depend on whether the case considered is
supersymmetric.) The prediction for
$m_s$ is
$209\, GeV$.

The $1\sigma$ experimental limits on the CKM parameter $\mid
{V_{ub}/V_{cb}}\mid$ are
\begin{eqnarray}
\left| {V_{ub}\over V_{cb}}\right| =0.09\pm 0.04\, .
\end{eqnarray} For our example, the prediction is
$\mid {V_{ub}/V_{cb}}\mid=0.0605\sqrt{{{m_u\over m_d}\over 0.6}{1.27\,
GeV\over m_c}}$. The possible range for $\mid {V_{ub}/V_{cb}}\mid$ is
shown in Fig. 1a. For this typical example, we can see that $\mid
{V_{ub}/V_{cb}}\mid$ varies from the lower end of acceptability 0.05 up
to about 0.0665.

For the CP violating parameter $J$, we find $J\cdot 10^5=3.0 \left(
{\mid V_{cb}\mid \over 0.05}\right)^2$ when $m_u/m_d=0.6$ and
$m_c=1.27\, GeV$. In Fig. 1b for the case of $m_u/m_d=0.6$ and
$m_c=1.27\, GeV$, we plot $J$ as a function of $\mid V_{cb}\mid$ for
values of $\mid V_{cb}\mid$ less than 0.053 and greater than 0.043,
which is the allowed range of
$\mid V_{cb}\mid$ within its $1\sigma$  experimental limits. The plot
shows that under these conditions $J\cdot 10^5$ can range from 2.2 to
3.4. In Fig. 2, we also plot $\cos{\phi}$ as a function of $\mid
{V_{ub}/V_{cb}}\mid$ over its predicted range. This plot is of course
also applicable to the non-SUSY cases to be dicussed.  The range of
$\cos{\phi}$ shown is from 0.14 to 0.30. The signifigance of
$\cos{\phi}$ for experiment is given in ref. \cite{CPviol}.

Next, we look at the predictions made for $m_t$ and $\tan{\beta}$. In
ref
\cite{ADHRS,Barger}, it was determined that each value of $m_t$ has two
values of $\tan{\beta}$ associated with it. Since each value of
$\tan{\beta}$ has only one value of $m_t$ and one value of $\mid
V_{cb}\mid$ associated with it, in Fig. 1c we plot $m_t$ vs.
$\tan{\beta}$ and in Fig. 1d we plot
$|V_{cb}|$ vs.
$\tan{\beta}$. Here, we plot the region described by $\tan{\beta}\leq
60$ and
$m_t\geq 125\, GeV$. As in ref. \cite{ADHRS,Barger}, for each value of
$\tan{\beta}$ we numerically integrate the RGE's from the scale
$\mu_t=180\, GeV$ for different values of
$m_t$ until we find one that gives $\lambda_{b_U}$ and
$\lambda_{\tau_U}$ to be within $.1\%$ of each other at the grand
unification scale $M_U$. From recent direct top searches\cite{direct},
$m_t^{pole}\geq 131\, GeV$
\cite{direct}. According to the analysis of the most recent electroweak
data
\cite{topmass}, $m_t^{pole}\leq 180\, GeV$. The figure shows that the
top mass is within these bounds only for some values of small
$\tan{\beta}$ and for large $\tan{\beta}\sim 60$.

As in ref.
\cite{Barger}, we also plot in Fig. 1e the grand unification scale
couplings
$A$ and
$D$ as a function of
$\tan{\beta}$. At about $\tan{\beta}=58$, we can see that $D=A$ for the
example $m_b=4.35\, GeV$. (For both of the other two examples graphed,
$D=A$ for some $\tan{\beta}$ a little greater than 60.) In ref
\cite{ADHRS,Barger} it was shown that one may use the unification scale
condition $D=A$ to decrease by one the number of inputs in the ansatze
and hence increase its number of predictions to 5 SM parameters and
$\tan{\beta}$.  With $D=A$ at
$M_U$,
$\mid V_{cb}\mid$ \cite{ADHRS,Barger} can now also be predicted.

Finally, we review work done on the neutrino sector and the possibility
of there being an ansatze to predict the neutrino masses and the
leptonic mixing angles. In ref.
\cite{DHRneutrino}, DHR propose the following ansatze for the neutrino
Dirac mass matrix and Majorana mass matrix respectively:
\begin{eqnarray} M_{\nu N}=\left( \matrix{{0}&{-3C}&{0}\cr
{-3C}&{0}&{-3\kappa B}\cr  {0}&{-3\kappa B}&{-3A}}
\right) \kappa \sin{\beta}\label{neutI}
\end{eqnarray} and
\begin{eqnarray} M_{N N}=\left( \matrix{{0}&{C}&{0}\cr {C}&{0}&{0}\cr
{0}&{0}&{A}}\right) V\, ,\label{neutII}
\end{eqnarray} where $V$ is the superheavy singlet VEV and $\kappa =1$
or
$-1/3$.  The low mass neutrino mass matrix is then of the form
\begin{eqnarray} M_{\nu \nu}=M_{\nu N}M_{N N}^{-1}M_{\nu N}^{T}\, .
\end{eqnarray} Then, just as in the quark sector, from bilinear
transformations
$M_E^{diag}=V_e^LM_EV_e^{R\dagger}$ and
$M_{\nu \nu}^{diag}=V_\nu^LM_{\nu \nu}V_\nu^{R\dagger}$ that
diagonalize the lepton mass matrices one finds the leptonic CKM matrix
$V'=V_\nu V_e^{L\dagger}$. DHR then find the following neutrino mass
ratios and mixing angles:
\begin{eqnarray} {m_{\nu_\tau}\over m_{\nu_\mu}}&=&{1\over
3\kappa^2}\left( {B\over A}\right)^{-2}\, ,\\ {m_{\nu_\mu}\over
m_{\nu_e}}&=&9 \kappa^4 {m_c
\eta_u \over m_u \eta_c}\, ,\\
\theta_{\mu \tau}&\simeq &-2\kappa {B\over A}\, ,\\
\theta_{e \mu}&\simeq &\left[ {m_e\over m_\mu}+{m_{\nu_e}\over
m_{\nu_\mu}}-2\sqrt{{m_e m_{\nu_e}\over m_\mu m_{\nu_\mu}}}\cos{\phi}
\right]^{1\over 2}\, ,\\
\theta_{e\tau}&\simeq &{2\over 3}\kappa \sqrt{m_e\over m_\mu}{B\over
A}\, ,
\end{eqnarray} in which $B/A=|V_{cb}(M_U)|$.

For our example with $\kappa=1$  and assuming $\tan{\beta}$ to be
small, we find the following:
\begin{eqnarray} {m_{\nu_\tau}\over m_{\nu_\mu}}&=&278 \, ,\\
{m_{\nu_\mu}\over m_{\nu_e}}&=& 3680\, ,\\
\sin^2{\theta_{\mu \tau}}&=&0.0191 \, ,\\
\sin^2{\theta_{e \mu}}&=&0.0177 \, ,\\
\sin^2{\theta_{e\tau}}&=& 1.03\times 10^{-5}\, ,
\end{eqnarray} where we have used $|V_{cb}|=0.05$, $m_u/m_d=0.43$ and
$m_c=1.23\, GeV$. We used $m_u/m_d=0.43$ and $m_c=1.23\, GeV$ to get
$\sin^2{\theta_{e \mu}}$ as low as possible. The value of
$\sin^2{\theta_{e
\mu}}$ and the mass ratios found in this example are to be compared
with the small mixing-angle non-adiabatic solution window ($\Delta
m^2\simeq (0.3-1.2)\times 10^{-5}eV^2$ and
$\sin^2{\theta_{e \mu}}\simeq (0.4-1.5)\times 10^{-2}$) which is in
agreement with all experimental data \cite{neutrino}.  The value of
$m_{\nu_\tau}$ is $\sim 1\, eV$. The $\kappa=-1/3$ scenario can only
provide neutrino masses and mixing that lie well between the small and
large angle
$90\%$ confidence limit MSW solution windows \cite{DHRneutrino}.

\section{Ansatze in SM} As discussed in Section 2, the unification of
$m_b$ and $m_\tau$ at high energies is not possible in the SM.  Wanting
both to have an acceptable value of $m_b$ and use mass matrices as
similar as possible to the GJ form, we will use the following ansatze
at the grand unification scale:
\begin{eqnarray} U\sim \left( \matrix{ {0}&{C}&{0}\cr {C}&{0}&{B}\cr
{0}&{B}&{A}\cr }\right)\, ,\, D\sim \left(
\matrix{ {0}&{F}&{0}\cr {F}&{E}&{0}\cr {0}&{0}&{ D+d}\cr }\right)\, ,\,
E\sim \left( \matrix{ {0}&{F}&{0}\cr {F}&{-3E}&{0}\cr {0}&{0}&{ D-3d}\cr
}\right)\, ,\label{SMunifmatrix}
\end{eqnarray} where $A$, $B$,$C$, $D$, $d$, $E$, and $F$ are complex
parameters, with $|A|>>|B|>>|C|$ and $|D+d|\sim |D-3d|>>|E|>>|F|$.

Below grand unification scale, the zero entrees in the mass matrices
will develop small finite values. However, we have found the values
that these entrees develop when one takes the energy scale from grand
unification scale down to the intermediate breaking scale are
negligible. So, it is a good approximation to take the ansatze at the
intermediate breaking scale. (Most importantly,
$|V_{cb}|/\sqrt{m_c\over m_t}$ does not evolve between $M_U$ and
$M_I$.) After rotating away all but one unavoidable phase
$\phi$ in the mass matrices by redefinition of the phases of the fermion
fields
\cite{DHR}, we take the ansatz at the intermediate breaking scale to be
\begin{eqnarray} U\sim \left( \matrix{ {0}&{C}&{0}\cr {C}&{0}&{B}\cr
{0}&{B}&{A}\cr }\right)\, ,\, D\sim \left(
\matrix{ {0}&{Fe^{i\phi}}&{0}\cr {Fe^{-i\phi}}&{E}&{0}\cr {0}&{0}&{\mid
D+d\mid}\cr }\right)\, ,\, E\sim \left( \matrix{ {0}&{F}&{0}\cr
{F}&{-3E}&{0}\cr {0}&{0}&{\mid D-3d\mid}\cr }\right)\, .\label{matrix}
\end{eqnarray}

Although this ansatze lacks bottom-tau Yukawa coupling unification, it
uses the same number, 3, of parameters to describe the third generation
masses as does the MSSM or 2HSM cases with bottom-tau Yukawa coupling
unification because they require the additional parameter
$\tan{\beta}={\kappa_u/ \kappa_d}$. $D$ and $d$ may always be chosen to
satisfy experimentally determined values of $m_b$ and $m_\tau$, but do
not make predictions. Besides the two parameters $D$ and $d$ our
ansatze has 6 other parmaters, and other than $m_b$ and $m_\tau$ the SM
has 11   fermion sector parameters. So, we can make 5 predictions from
the 6 of these 11  fermion sector parameters that are best determined.
We use $m_e$, $m_\mu$,
$m_c$, $m_u/m_d$, $\mid V_{cb}\mid$,  and $\mid V_{us} \mid$ as inputs.
In the last section, we quoted acceptable values for these parameters.

Now, we look at the predictions for
$m_t$, $m_s$, $m_s/ m_d$, $\mid {V_{ub}\over V_{cb}} \mid$, and the CP
violation parameter $J$ (or $\cos{\phi}$). Note, these are the same SM
quantities as predicted for the DHR model without top-bottom Yukawa
coupling unification. (The DHR model predicts these 5 SM parameters and
also the SUSY parameter $\tan{\beta}={\kappa_u/ \kappa_d}$.) We will
look at predictions for  two cases. For case (a) we use
$M_I=10^{10.94}\, GeV$,  and for case (b) we use $M_I=10^{14}\, GeV$.

First, from Eqs\@. (\ref{mt}), (\ref{mc}), (\ref{vcb}), and (\ref{Ku})
and
\begin{eqnarray} m_t={m_c/\eta_c \over
\sqrt{|V_{cb}|^4+{K_u\over \kappa^2 A_u^2}\left( {m_c\over
\eta_c}\right)^2}}\, .
\end{eqnarray}  We show running mass $m_t$ vs.
$\mid V_{cb}\mid$ for the SM scenario in Fig. 3a for case (a) and in
Fig. 4a for case (b). In case (a) we see that
$\mid V_{cb}\mid$ can be as low as 0.039, and in case (b) $\mid
V_{cb}\mid$ can be as low as 0.037 for running mass $m_t$ less than
$200\, GeV$. For
$\mid V_{cb}\mid$ within its
$1\sigma$  limits, in case (a) $m_t$ can be as low as $145\, GeV$ and
in case (b) $m_t$ can be as low as $140\, GeV$.

Now, we look at the other 4 predictions. These 4 predictions all take
the same form as in the original DHR ansatze and are given by Eq\@.
(\ref{i}), Eq\@. (\ref{ii}), Eq\@. (\ref{iii}), and Eq\@. (\ref{iv}).
Of course, the prediction for $m_s/m_d$ is the same as before
${m_s/ m_d}=24.71$ because it only depends on the ratio $m_e/m_\mu$. The
other three predictions are proportional to  the ratio of the gauge
contribution for the down quark masses to the gauge contribution for the
charged lepton masses $R_{d\over e}={A_d/ A_e}$.

Since the prediction for $m_s$ is proportional to $R_{d\over e}$,   the
range of predicted values of $m_s$ in the SM case (a) has to be lower
than the range of predicted value in the SM case (b). In case (a) we
find
\begin{eqnarray} m_s=166_{-21}^{+29}\, GeV
\end{eqnarray} , and in case (b) we find
\begin{eqnarray} m_s=184_{-23}^{+33}\,GeV\, .
\end{eqnarray} The uncertainties that we give are due to the
uncertainty in
$\alpha_3(M_Z)$. The value in our MSSM example was $m_s=209\, GeV$,
which is contained in the upper part of the range of values for the SM
case (a).

Also, the prediction for $|{V_{ub}\over V_{cb}}|$ is proportional to
$R_{d\over e}$. So, once again, we expect that the range of predicted
values for $| {V_{ub}\over V_{cb}}|$  in the SM case (a) to be  lower
than the range of predicted values in the SM case (b). In the SM case
(a) we find
\begin{eqnarray}
\left| {V_{ub}\over  V_{cb}}\right|
=(0.054^{+0.004}_{-0.003})\sqrt{{{m_u\over m_d}\over 0.6}{1.27\,
GeV\over m_c}}\, ,
\end{eqnarray} and in the SM case (b) we find
\begin{eqnarray}
\mid {V_{ub}\over V_{cb}}\mid =(0.057^{+0.004}_{-0.003})\sqrt{{{m_u\over
m_d}\over 0.6}{1.27\, GeV\over m_c}}\, .
\end{eqnarray} The uncertainties given here are  due to the uncertainty
in
$\alpha_3(M_Z)$. The value in our MSSM example was
$\left| {V_{ub}\over V_{cb}}\right| =0.065\sqrt{{{m_u\over m_d}\over
0.6}{1.27\, GeV\over m_c}}$, which is contained in the upper part of the
range of values for the SM case (a). We show the range of good values
for
$\mid {V_{ub}\over V_{cb}}\mid$  in Fig\@. (\ref{sma}) for the SM case
(a) and in Fig\@. \ref{smb} for the SM case (b).

Being proportional to $R_{d\over e}$, one expects the CP violation
parameter
$J$ to have a lower range of predicted values in the SM case (a) than
in the SM case (b). When $m_u/m_d=0.6$ and $m_c=1.27\, GeV$, we find
\begin{eqnarray} J\cdot 10^5=(2.6^{+0.3}_{-0.2})\left( {| V_{cb}| \over
0.05}\right)^2
\end{eqnarray} for the SM case (a) and
\begin{eqnarray} J\cdot 10^5=(2.8\pm 0.2)\left( {| V_{cb}| \over
0.05}\right)^2
\end{eqnarray} for the SM case (b). This is to be compared with
$J\cdot 10^5=3.0\left( {| V_{cb}| \over 0.05}\right)^2$ in the MSSM
case. The prediction for case (a) is ploted in Fig. 3c, and the
prediction for case (b) is plotted in Fig. 4c. The predicted values for
$\cos{\phi}$ can again be found from Fig. 2 for the predicted ranges of
$| {V_{ub}/ V_{cb}}|$.

To complete this section, we will consider neutrino mass matrices of the
form  given in Eq\@. (\ref{neutI}) and  Eq\@. (\ref{neutII}). However,
as a good approximation we will take the matrices at $M_I$ instead of
$M_U$. Following the same analysis as discussed in the last section, we
find the following for case (a) when
$|V_{cb}|=0.05$,
$m_u/m_d=.51$ ,
$m_c=1.27\, GeV$, and $\alpha_{3c}(M_Z)=0.118$:
\begin{eqnarray} {m_{\nu_\tau}\over m_{\nu_\mu}}=109 \, ,\\
{m_{\nu_\mu}\over m_{\nu_e}}= 3720\, ,\\
\sin{\theta_{\mu \tau}}^2=0.0483 \, ,\\
\sin{\theta_{e \mu}}^2=0.0176 \, ,\\
\sin{\theta_{e\tau}}^2= 2.64\times 10^{-5}\, ,
\end{eqnarray} and we find the following for case (b) when
$|V_{cb}|=0.05$,
$m_u/m_d=.46$ , $m_c=1.27\, GeV$, and $\alpha_{3c}(M_Z)=0.118$:
\begin{eqnarray} {m_{\nu_\tau}\over m_{\nu_\mu}}=106 \, ,\\
{m_{\nu_\mu}\over m_{\nu_e}}= 3730\, ,\\
\sin^2{\theta_{\mu \tau}}=0.0493 \, ,\\
\sin^2{\theta_{e \mu}}=0.0176 \, ,\\
\sin^2{\theta_{e\tau}}= 2.69\times 10^{-5}\, .
\end{eqnarray} Because $|V_{cb}|$ becomes larger at higher energies in
the SM whereas it becomes smaller at higher energies in the MSSM, the
values for
$\sin^2{\theta_{e \mu}}$ are virtually the same in the MSSM and SM cases
whereas the ratio ${m_{\nu_\tau}/ m_{\nu_\mu}}$ is more than twice as
big in the MSSM example than in the SM cases. The value of
$m_{\nu_\tau}$ is $\sim {1\over 2}\, eV$.

\section{Ansatze in 2HSM} For the 2HSM case, we first use use an
ansatze of the form given in Eq\@. (\ref{GJ}) at grand unification
scale. Although the zero entrees in the Yukawa matrices will develop
relatively small values between $M_U$ and $M_I$,
$|V_{cb}|/{\sqrt{m_c\over m_t}}$ does not evolve over that range and so
as a  good approximation one can effectively take the ansatze at $M_I$
in the form of Eq\@. (\ref{rotatedaway}).    As does the DHR ansatze,
this ansatze has 8 parameters. So, it is possible to predict 5 SM
fermion sector parameters and the 2HSM parameter
$\tan{\beta}$ in terms of the 8 best measured SM fermion sector
parameters. Of course, we choose the same 5 input parameters as in
Section 3. The expressions for the 4 output parameters $m_s$,
$m_s/ m_d$, $\mid {V_{ub}\over V_{cb}} \mid$, and the CP violation
parameter
$J$ (or $\cos{\phi}$)  again are given by Eq\@. (\ref{i}), Eq\@.
(\ref{ii}), Eq\@. (\ref{iii}), and Eq\@. (\ref{iv}). Since in the 2HSM
$R_{d\over e}={A_d/ A_e}$ has values within a few percent of its values
in the SM case (a), these 4 2HSM case predictions will only be slightly
different than the predictions of these 4 parameters that were given
for the SM case (a). Those predictions are already given in Table 2 and
Fig. 3. However, we do need to discuss the predictions for $m_t$ and
$\tan{\beta}$.

If we are to require $\lambda_{b_U}=\lambda_{\tau_U}$ but not
$\lambda_{t_U}=\lambda_{\tau_U}$, then we must have two Higgs
biodoublets  instead of one in the intermediate scale effective theory.
(Hence for this case the model needs two complex ${\bf 10}$'s instead
of the minimal one complex ${\bf 10}$.) One Higgs doublet from each of
these bidoublets is then assumed to contain a VEV and appear in the
2HSM effective theory below
$M_I$. (One Higgs doublet is $\phi_u$ and the other is $\phi_d$.) For
the more interesting case of
$\lambda_{t_U}=\lambda_{b_U}=\lambda_{\tau_U}$, the model only needs one
Higgs bidoublet appearing at intermediate scales, and hence the model
only needs the minimal one complex ${\bf 10}$ Higgs field. The
$A_\alpha$'s and the
$K_u$'s which we give in Table 1 for the 2HSM case and use in this
section were calculated for the assumption of only one Higgs bidoublet
having a mass less than $M_U$. The $M_I$ we use is calculated according
to the principal of minimal fine-tuning and for when
$\alpha_{3c}(M_Z)=0.018$. The values of the
$A_\alpha$'s and the
$K_u$'s that are calculated for the 2 Higgs bidoublet case are similar
to the corresponding values given for the single Higgs bidoublet case,
and one would expect these differences to be smaller than the
uncertainties in the
$A_\alpha$'s and the
$K_u$'s due to possible threshold corrections which we ignore for the
sake of simplicity.

When the assumption of $\tan{\beta}$ being small is made, $m_t$ and
$\tan{\beta}$ may be predicted to a very good approximation by the
following equations:
\begin{eqnarray} m_t(m_t)&=&{m_c/\eta_c \over \mid V_{cb}\mid^2}{b\over
\tau}\, ,\\
\sin{\beta}&=&{\sqrt{K_u}\over A_u \kappa}{m_c/\eta_c \over \mid
V_{cb}\mid^2}
\left( {\tau \over b}\right)^{8} \left[ \left( {\tau \over
b}\right)^{18}-1\right]^{-{1\over 2}}\, ,
\end{eqnarray} and for the intermediate breaking scale top quark Yukawa
coupling
\begin{eqnarray} A=K_u^{-{1\over 2}}\sqrt{\left( {\tau \over
b}\right)^{18}-1}
\end{eqnarray}    where we have again used
$\tau ={m_\tau \over \eta_\tau A_e}$ and
$b ={m_b \over \eta_b A_d}$,    and $m_t$ is the top quark running
mass.       In order to investigate the situation for when
$\tan{\beta}$ is not small we must numerically integrate the Yukawa
RGE's to find for each value of
$\tan{\beta}$ a value of $m_t$ for which $\lambda_{b_I}$ agrees with
$\lambda_{\tau_I}$ to within $0.1\%$.

We have found two seperate ranges of $\tan{\beta}$ that give values for
the running mass $m_t$ between $125\, GeV$ and $200\, GeV$. One region
is for
$\tan{\beta}\sim 1$ and has $A$ much greater than $D$. In the other
region,
$\tan{\beta}$ is greater than about 55 and $D$ is of the same order as
or larger than $A$. It is not surprising that we find two separate
regions in
$\tan{\beta}$. One expects the $m_t$ vs. $\tan{\beta}$ plots for the
2HSM case to have the same shape as the $m_t$ vs. $\tan{\beta}$ plot
for the MSSM case in Fig. 1c, but one also expects as discussed in
Section 2 that in both cases when $A$ is much larger than $D$ and
$\sin{\beta}\approx 1$ the top mass required by the $M_I$ scale
condition
$m_b=m_\tau$ will be close to
$\kappa A_u /\sqrt{K_u}$.  While $\kappa A_u /\sqrt{K_u}$ is a little
smaller than
$200\, GeV$ in the MSSM case, it is larger than $200\, GeV$ in the 2HSM
case. Hence, one would expect $m_t$ to be unacceptably large for
intermediate values of $\tan{\beta}$ for which $\sin{\beta}\approx 1$
and
$A$ is much larger than $D$.

For the case that
$\alpha_3(M_Z)=0.111$ and $m_c=1.22\, GeV$, we find for  a small span of
$\tan{\beta}$ ($\sim 1$) from about $0.6$ to about $1.7$, the running
mass
$m_t$ takes values from
$125\, GeV$ to $200\, GeV$. Within this region, $\mid V_{cb}\mid$
could be as low as about 0.0515.  When $m_b$ has the values $4.35\,
GeV$,
$4.25\, GeV$, and
$4.15\, GeV$, the $M_I$ scale coupling $A=\lambda_{t_I}$ has the values
1.4, 1.8, and 2.3, respectively. (Larger input values for $m_b$ give
smaller values for $A$.) However, from Eq\@. (\ref{lambdatI}) we find
that
$A=\lambda_{t_I}$ can have a maximum value of 1.26. The effect of using
larger values of
$\alpha_3(M_Z)$ is to require larger values of $A$ than just given for
the
$\alpha_3(M_Z)=0.111$ case. (e.g. When $m_b=4.35\, GeV$ and
$\alpha_3(M_Z)=0.118$, $A$ must be 2.3.) This lower region is ruled out
in the  scheme we are using unless the running mass
$m_b$ is larger than about
$4.4\, GeV$ and $\alpha_{3c}(M_Z)$ is near its lower end of
acceptability.

In Fig. 5a and Fig. 5b, we show the running mass $m_t$ vs.
$\tan{\beta}$ and $\mid V_{cb}\mid$ vs. $\tan{\beta}$ respectively for
the higher region of $\tan{\beta}$   for the case that
$\alpha_3(M_Z)=0.111$ and $m_c=1.22\, GeV$. In the $m_t$ vs.
$\tan{\beta}$ plot, we  plot $m_t$ for values of $M_I$ scale Yukawa
couplings $A$ and $D$ less than 1.3. We see that for  $m_b=4.35\, GeV$,
$m_t$ can be as low as
$150\, GeV$. In the
$\mid V_{cb}\mid$ vs. $\tan{\beta}$ plot, we can see that $\mid
V_{cb}\mid$ is never within the $1\sigma$ limits of $\mid V_{cb}\mid$
but can be within its $90\%$ confidence limits. In Fig. 5c, we also
show the unification scale couplings $A$ and $D$ as a function of
$\tan{\beta}$. We can see that for the case with $m_b=4.35\, GeV$
top-bottom-tau unification ($D=A$) is possible for $A\approx 0.8$.

In Fig. 6a through Fig. 6d, we show $m_b$, $m_t$, $\mid V_{cb}\mid$, and
$\tan{\beta}$ as a function of $A$ when $D=A$ for the case where
$\alpha_3(M_Z)=0.111$ and $m_c=1.22\, GeV$. Using a value of $m_b$ as an
input determines a value for $A$, but only values of $m_b$ more than
$4.25\, GeV$ predict values of $m_t$ less than $200\, GeV$. In fact,
for $m_b\leq 4.4\, GeV$ the top running mass is predicted to be high,
greater than $180\, GeV$. Once again, the possible range for $\mid
V_{cb}\mid$ lies outside of its $1\sigma$ limits but within its
$90\%$ confidence limits. The value for $\tan{\beta}$ is predicted to be
between 57.5 and 65 for $m_t<200\, GeV$. The $M_I$ scale Yukawa
coupling $A$ takes values from 0.73 to 1.00 for $m_b\leq 4.4\, GeV$.

Fig. 6a through Fig. 6d for the 2HSM case can be compared with the
situation in the MSSM. In Fig. 7a through Fig. 7d, we show $m_b$,
$m_t$, $\mid V_{cb}\mid$, and
$\tan{\beta}$ as a function of $A$ when $D=A$ for the case when
$\alpha_3(M_Z)=0.121$, $M_S=180\, GeV\, GeV$ and $m_c=1.22\, GeV$. We
see that in the MSSM, having $m_b$ within the $90\%$ limits given in
ref.
\cite{experiment} correspond to lower values of $m_t$ than in the 2HSM
case just discussed. For example, $m_b=4.4\, GeV$ corresponds to a
running mass
$m_t=174.5\, GeV$, which is a pole mass of $183\, GeV$. Although its
values are found to be lower than in the 2HSM,
$|V_{cb}|$ comes out just above its $1\sigma$ limits. As in the 2HSM
case,
$\tan{\beta}\sim 60$.

Bottom-tau Yukawa coupling unification in the 2HSM with
$\alpha_3(M_Z)=0.118$ requires high values of $m_b$ to keep both of the
couplings $A$ and
$D$  from being too large. For example when
$m_b=4.4\, GeV$, $D$ can only be as small as 2.03 when $A=1.02$,
$m_t=200\, GeV$,
$\tan{\beta}=74.9$, and $\mid V_{cb}\mid$ is 0.054 for $m_c=1.22\,
GeV$. A similar problem results if we increase $M_I$. We find that the
unification of the bottom and tau Yukawa couplings is only feasible in
the 2HSM when
$M_I<<M_U$ and $\alpha_{3c}(M_Z)$ is low, near 0.111.

\section{$U(1)^3$ symmetry and induced VEV's to give mass matrices}

Recently the authors of ref. \cite{BM} have shown that if certain
reasonable  assumptions are made then the neutrino mass ratios and
leptonic mixing angles are completely determined by the 13 SM fermion
sector parameters within the context of minimal SO(10)
grandunification. Their 13 parameter  model is capable of generating
all of the fermion masses and quark mixing angles and predicting the
neutrino spectrum without  depending upon any flavor symmetries.
Crucial to their scheme is the observation that the electroweak
breaking VEV of the {\bf 10} representation Higgs field will induce a
small VEV in the super heavy bidoublet of the {\bf 126} representation
Higgs field. Their model of course has little predictive ability in the
SM sector.

In this section we give an example of a scheme that makes use of the
idea of induced VEV's from super heavy fields, but at the same time
limiting the structure of the mass matrices by using softly broken
global symmetries. Specifically, we use
$U(1)^3$ symmetry to generate mass matrices similar to Eq\@.
(\ref{SMunifmatrix}) which account for the hierarchy of masses and
mixing angles. We shall have to go beyond the minimal SO(10) model to
accomplish this.

We consider the possibility  that  SO(10) gauge symmetry is broken to
the gauge symmetry
$2_L\, 2_R\, 4_C$ by a {\bf 210} representation Higgs field. At the next
stage, symmetry is broken to
$2_L\, 2_R\, 1_{B-L}\, 3_c$ by  {\bf 210} as well as  a {\bf 45}
representation of Higgs field. Breaking to to the SM is done by a {\bf
126} representation, and then finally the electroweak symmetry is
broken by a complex {\bf 10} representation. In our example, we find
that we need two  super heavy {\bf 10} representations and two  super
heavy {\bf 126} fields. The super heavy fields have only very small
induced VEV's. The {\bf 10} representation  that does the electroweak
symmetry breaking we will denote by
${\bf 10_3}$, and the {\bf 126} representation Higgs field that breaks
the symmetry $2_L\, 2_R\, 4_C$ to $2_L\, 2_R\, 1_{B-L}\, 3_c$ we will
denote by
${\bf 126_3}$. We show in Table 3 all the fields that we employ and
their transformation properties under three different U(1) symmetries
$U(1)_X$,
$U(1)_Y$, and $U(1)_Z$. All bidoublets are super heavy except that of
the
${\bf 10_3}$ field. The operators that give the fermion masses are
shown in Fig. 8. These operators give the following Yukawa matrices:
\begin{eqnarray}  U&= &\left(
\matrix{ {0}&{C}&{0}\cr {C}&{E}&{B}\cr {0}&{B}&{A+a}\cr }\right)\, ,\,
D=
\left( \matrix{ {0}&{C r_C}&{0}\cr {C r_C}&{E r_E}&{B r_B}\cr {0}&{B
r_B}&{A r_A+ar_a}\cr }\right)\, ,\nonumber\\E&= & \left( \matrix{
{0}&{C r_C}&{0}\cr {C r_C}&{-3E r_E}&{B r_B}\cr {0}&{B r_B}&{A
r_A-3ar_a}\cr }\right)\, ,\label{inducedmatrix}
\end{eqnarray}  where the $r_i$'s are ratios of the ``down" VEV's to the
``up" VEV's in the operators. These Yukawa matrices go to those of our
SM case in the limit of small $r_B$ and $r_E$ large compared to 3.

It is pointed out in ref. \cite{RRRso10} that a four-fold symmetrized
product of the {\bf 126}-dimensional representation is an SO(10)
singlet. Hence terms in the Lagrangian such as $\lambda ({\bf
126_i})^4_S$ will explicitly break a U(1) symmetry to discrete symmetry
if ${\bf 126_i}$ has a U(1) charge. We can use the term
$\lambda ({\bf 126_1})^4_S$ to break U(1) quantum numbers X, Y, and Z
to a mod 8, a mod 16, and a mod 8 discrete symmetry respectively and
avoid massless Nambu-Goldstone bosons.

We note that in this scheme one can not determine the neutrino sector
without making further assumptions. However, we still should check to
see if the scheme is capable of generating low mass neutrinos and
leptonic mixing angles that are in a range to provide an explanation
for the observed solar neutrino deficit via neutrino oscillation. Our
scheme provides a Majorana mass matrix with 3 unknown couplings to the
three ${\bf \overline{126}}$ representation Higgs fields and which is
of the form
\begin{eqnarray} M_{NN}=
\left( \matrix{ {\beta}&{0}&{0}\cr {0}&{\alpha}&{0}\cr {0}&{0}&{1}\cr
}\right)V\, ,
\end{eqnarray} where $V\sim M_R$ and
$\alpha$ and $\beta$ are in general complex and may be assumed to be
small.  We assume the
$(1,3,10)$ submultiplets, given in
$2_L\, 2_R\, 4_C$ notation, of the fields ${\bf 126_2}$ and ${\bf
126_1}$ have masses near the unification scale, and that they acquire
small VEV's. We do not explain these small VEV's, but we note that they
could result from a more complicated Higgs structure. The neutrino
Dirac mass matrix at $M_U$ is approximately the same as ${\bf
U}\kappa$. We find that it is possible to get the neutrino spectrum
into the previously mentioned small-angle adiabatic solution window,
$\Delta m^2\simeq (0.3-1.2)\times 10^{-5}eV^2$ and
$\sin^2{\theta_{e \mu}}\simeq (0.4-1.5)\times 10^{-2}$, when $|\alpha
|<<1$ and
$|\beta |<<|\alpha |$ provided we give phases to the SM singlet VEV's.
For example, if we assume phases our zero and use $\alpha =0.005$ and
$\beta =\alpha^2$ for when
$|V_{cb}|=0.05$ we get
$m_{\nu_\tau}/m_{\nu_\mu}\approx 500$ and
$\sin^2{\theta_{e \mu}}\approx 0.018$. However for example , if we give
a complex phase of $\phi$, $2\phi$ and 0 to the third, second and first
generation diagonal entries in the Majorana mass matrix , then for
$|V_{cb}|=0.05$ we get
$m_{\nu_\tau}/m_{\nu_\mu}\approx 750$ and
$\sin^2{\theta_{e \mu}}\approx 0.01$, which is an acceptable solution
to the solar neutrino problem.

\section{Summary and Conclusions} In this paper, we have examined the
predictive ability of fermion mass ansatzes in non-SUSY SO(10) grand
unification in contrast to SUSY SO(10) since there is still no direct
evidence for SUSY.  We have considered the two possibilities that
between the scale of the top mass and the scale $M_I$ the effective
theory is  the SM and that it is the 2HSM. We have compared these cases
to the case where between the scale of the top mass and $M_U$ the
effective theory is the MSSM, where the maximal SM parameter predictive
ability is six parameters with
$|V_{cb}|$ a little large or 5 parameters all within $1\sigma$
experimental  limits. We have not considered ansatzes such as given in
ref. \cite{ADHRSII} where certain  relations are assumed between all of
the entrees of the up and down quark Yukawa matrices with the result of
the predictive ability being improved.

In the SM case, we find the condition $m_b=m_\tau$ at the unification
scale
$M_U$ is impossible to maintain with $m_t^{pole} \geq 130\, GeV$ and
$m_b<5\, GeV$. Nevertheless, we are able to predict 5 SM parameters to
be within their
$1\sigma$ experimental limits. Specifically, $m_t$ is in the range of
about
$150\, GeV$ to  $180\, GeV$ for $|V_{cb}|$ in the upper half of its
$1\sigma$ range. This is shown in Fig. 3a and Fig. 4a for the case of
$M_I\sim 10^{11}\, GeV$ and
$M_I\sim 10^{14}\, GeV$ respectively. The results for the MSSM are quite
similar for the ranges of $m_t$ and $|V_{cb}|$ that are permissable. The
values of $|V_{ub}/V_{cb}|$, $m_s$, and $J$ for the SM and the MSSM
cases are shown in Table 2. As can be seen they are quite similar and
lie  within the $1\sigma$ experimental limits. These 3 parameters are
found to depend somewhat on the scale that the Pati-Salam group is
broken at. The predictions for these 3 parameters increase when the
intermediate scale
$M_I$ is increased. In all cases
$|V_{ub}/V_{cb}|$ is seen be on the lower end of its acceptable range.
For the SM case with $M_I\sim 10^{11}\, GeV$ $|V_{ub}/V_{cb}|$ must be
less than about 0.064, while in the SM case with $M_I\sim 10^{14}\,
GeV$ it can be as high as about 0.068. As usual, the prediction for
$m_s/m_d$ only depends on
$m_\mu/m_e$ and is found to be 24.73, within experimental bounds.

As in the MSSM and unlike in the SM, in the 2HSM both $m_b=m_\tau$ and
with large $\tan{\beta}$  unification of the top, bottom and tau Yukawa
couplings at the gauge unification scale  are possible. We find we can
predict
$\tan{\beta}$ and 6 SM parameters for the case where the top, bottom
and tau Yukawa couplings are unified at high energies. This is found
only to work when
$\alpha_{3c}(M_Z)$ is near 0.111, and so could be ruled out with better
experimental determination of $\alpha_{3c}(M_Z)$.  The predictions for
the 4 parameters
$m_s/m_d$,
$|V_{ub}/V_{cb}|$, $m_s$ and
$J$ are essentially the same as for the SM. However, as shown in Fig. 6a
$|V_{cb}|$ is predicted to be above its $1\sigma$ limits. In fact, only
for
$m_t$ above $180\, GeV$ is $|V_{cb}|$ within its $90\%$ confidence
limits. Of course, by adding another parameter to the ansatze and
decreasing its its number of predictions by one
$|V_{cb}|$ may be allowed to be in its $1\sigma$ range. However, from
comparison of Fig. 6a and Fig. 6b one can see that for $m_t$ to be less
than
$180\, GeV$, the running mass $m_b$ must be greater than $4.4\, GeV$.
On the other hand, if we give up the unification of the top and bottom
Yukawa couplings but retain $m_b=m_\tau$ above $M_I$, then it is
possible for the top pole mass to be below $180\, GeV$. In this case,
$|V_{cb}|$ lies above its
$1\sigma$ limits but  within its $90\%$ confidence limits. The
predictions for $m_s/m_d$,
$|V_{ub}/V_{cb}|$, $m_s$ and
$J$ are essentially unchanged.

This work has been supported by the Department of Energy, Grant No.
DE-FG06-85ER-40224.

\vfill

\leftline{{\Large\bf Figure  captions}}

\begin{itemize}
\item[Fig. 1~:] {{In this figure, we show the following results for the
MSSM example discussed in Section 3. {\bf Fig. 1a}: The prediction for
$|V_{ub}/V_{cb}|$ vs. input values of $m_u/m_d$. The short dashed line,
solid line and the long dashed line represent the cases where $m_c$ is
$1.22\, GeV$, $1.27\, GeV$ and $1.32\, GeV$ respectively. {\bf Fig.
1b}: The prediction for the CP violation parameter
$J$ vs. input values of $|V_{cb}|$. The short dashed line, solid line
and the long dashed line represent the cases where $m_c$ is $1.22\,
GeV$, $1.27\, GeV$ and $1.32\, GeV$ respectively. {\bf Fig. 1c}: The
prediction for the running mass $m_t$ as a function of
$\tan{\beta}$ for $m_t>125\, GeV$ and $\tan{\beta}\leq 60$.   The short
dashed line, solid line and the long dashed line represent the cases
where
$m_b$ is $4.35\, GeV$,
$4.25\, GeV$ and $4.15\, GeV$ respectively.    {\bf Fig. 1d}: The CKM
matrix parameter $| V_{cb}|$ as a function of $\tan{\beta}$. The long
dashed line, the solid line and the short dashed line represent the
same values of $m_b$ as in Fig. 1c and also the values $1.22\, GeV$,
$1.27\, GeV$ and $1.32\, GeV$ for $m_c$ respectively. {\bf Fig. 1e}:
The
$M_U$ scale top and bottom Yukawa couplings as a function of
$\tan{\beta}$. The long dashed line, the solid line and the short dashed
line represent the same values of $m_b$ as in Fig. 1c.}\label{mssm}}

\item[Fig. 2~:] {{The cosine of the complex phase that appears in the
DHR ansatze as a function of the input
$|V_{ub}/V_{cb}|$.}\label{cosphi}}

\item[Fig. 3~:] {{In this figure we show the following predictions for
the SM case (a) with $M_I=10^{10.93}\, GeV$ discussed in Section 4.
{\bf Fig. 3a}: The prediction for the running mass $m_t$ vs.
$|V_{cb}|$. The short dashed line represents the case where $m_c=1.22\,
GeV$ and
$\alpha_{3c}(M_Z)=0.125$. The solid line represents the case where
$m_c=1.27\, GeV$ and
$\alpha_{3c}(M_Z)=0.118$. The long dashed line represents the case where
$m_c=1.32\, GeV$ and
$\alpha_{3c}(M_Z)=0.111$. {\bf Fig. 3b}: The prediction for
$|V_{ub}/V_{cb}|$ vs. input values of $m_u/m_d$.  The long dashed line,
the solid line and the short dashed line represent the same values of
$m_c$ and
$\alpha_{3c}(M_Z)$ as in Fig. 3a. {\bf Fig. 3c}: The prediction for the
CP violation parameter $J$ vs. input values of $|V_{cb}|$.  The long
dashed line, the solid line and the short dashed line represent the
same values of
$m_c$ and
$\alpha_{3c}(M_Z)$ as in Fig. 3a.}\label{sma}}

\item[Fig. 4~:] {{In this figure we show some predictions for the SM
case (b) with $M_I=10^{14}\, GeV$ discussed in Section 4. Fig. 4a, Fig.
4b and Fig. 4c are described by the captions for Fig. 3a, Fig. 3b and
Fig. 3c respectively.}\label{smb}}

\item[Fig. 5~:] {{In this figure we show the following predictions for
the 2HSM case with
$M_I=10^{11.28}\, GeV$ discussed in Section 4. {\bf Fig. 5a}: The
running mass $m_t$ vs. $\tan{\beta}$ with the dashed line and the solid
line representing $m_b=4.35\, GeV$ and $m_b=4.25\, GeV$ respectively.
We show
$m_t$ between $125\, GeV$ and
$200\, geV$. {\bf Fig. 5b}: The CKM parameter $|V_{cb}|$ as a function
of
$\tan{\beta}$ with $m_c=1.22\, GeV$ and the dashed and solid line being
representing the same as in Fig. 5a. {\bf Fig. 5c}: The $M_I$ scale top
and bottom Yukawa couplings $A$ and $D$ plotted as a function of
$\tan{\beta}$ with the dashed and solid lines representing the same as
in Fig. 5a.}\label{twoHSM}}

\item[Fig. 6~:] {{For the case of $A\equiv
\lambda_{t_I}=\lambda_{b_I}=\lambda_{\tau_I}$ in the 2HSM case of
Section 5 with
$M_I=10^{11.28}\, GeV$, we plot running mass $m_b$, running mass $m_t$,
$|V_{cb}|$ and
$\tan{\beta}$ as a function of $A$ in Fig. 6a, 6b, 6c and 6d
respectively. In Fig. 6c, we use $m_c=1.22\, GeV$.}\label{twoHSMunif}}

\item[Fig. 7~:] {{For the case of $A\equiv
\lambda_{t_I}=\lambda_{b_I}=\lambda_{\tau_I}$ in the MSSM  with
$M_S=180\, GeV$, $\alpha_{3c}(M_Z)$ and threshold corrections having
been ignored for simplicity, we plot running mass
$m_b$, running mass
$m_t$,
$|V_{cb}|$ and
$\tan{\beta}$ as a function of $A$ in Fig. 7a, 7b, 7c and 7d
respectively. In Fig. 6c, we use $m_c=1.22\, GeV$.}\label{mssmunif}}

\item[Fig. 8~:] {{In this figure we show the operators discussed in
Section 6 that give the Yukawa couplings of Eq\@. (\ref{inducedmatrix})
from the fields given in Table 3.}\label{massopers}}

\end{itemize}

\vfill

\begin{table} \begin{center}
\begin{tabular}{|c||c|c|c|c|c|} \hline Scenario & $A_u$ & $A_d$ & $A_e$
&
$K_u$ & $R_{d\over e}={A_d\over A_e}$ \\ & & & & &\\ \hline \hline SM
case (a) &
$2.27\pm 0.05$ & $2.23\pm 0.05$ & 1.19 & $2.51\pm 0.05$ &
$1.87\pm 0.05$ \\ \hline SM case (b) & $2.69\pm 0.06$ & $2.62\pm 0.06$ &
1.26 & $3.98\pm 0.08$ &
$2.08\pm 0.04$ \\ \hline 2HSM & $2.32\pm 0.05$ & $2.28\pm 0.05$ & 1.20 &
$2.66\pm 0.05$ &
$1.90\pm 0.04$ \\ \hline MSSM & 3.45 & 3.36 & 1.50 & 9.55 & 2.24 \\
\hline
\end{tabular} \end{center} \caption[]{{\small\sf In this table, we show
the gauge contribution factors $A_\alpha $, defined in Eq\@.
(\ref{gaugecont}), the quantity $K_u$, defined in Eq\@. (\ref{K}) and
the ratio $R_{d\over e}={A_d\over A_e}$. In the first three cases
listed, we assume that the SO(10) grand unified group breaks to the
gauge group
$2_L\, 2_R\, 4_C$ at the scale $M_U$, and then the gauge symmetry $2_L\,
2_R\, 4_C$ is broken to either the SM or the 2HSM at the scale $M_I$.
In the SM case (a), the SM case (b) and the 2HSM case, we have assumed
$M_I=10^{10.93}\, GeV$, $M_I=10^{14}\, GeV$ and
$M_I=10^{11.28}\, GeV$ respectively. For the purpose of comparison, we
also give the results for the MSSM with the assumptions of gauge
coupling unification (for which we ignore threshold effects) and
$m_t\approx M_S=180\, GeV$  used to determine
$\alpha_{3c}(M_Z)=0.121$. } \label{gaugefactors}}
\end{table}

\begin{table} \begin{center} \begin{tabular}{|c||c|c|c|} \hline
Parameter & Prediction for & Prediction for & Prediction for\\  & SM
case(a) & SM case (b) & MSSM\\ \hline \hline
$m_s(1\, GeV)$ & $166^{+29}_{-21}\, GeV$ & $184^{+33}_{-23}\, GeV$ &
$209\, GeV$\\ \hline & & & \\ $\mid {V_{ub}\over V_{cb}}\mid$ &
$(0.054^{+0.004}_{-0.003})\sqrt{{{m_u\over m_d}\over 0.6}{1.27\,
GeV\over m_c}}$ & $(0.057^{+0.004}_{-0.003})\sqrt{{{m_u\over m_d}\over
0.6}{1.27\, GeV\over m_c}}$ & $0.0605 \sqrt{{{m_u\over m_d}\over
0.6}{1.27\, GeV\over m_c}}$\\ & & &\\ \hline $J\cdot 10^5$ & & &\\ for
${m_u\over m_d}=0.6$ &
$(2.6\pm 0.2) \left({\mid V_{cb}\mid \over 0.05}\right)^2$ &
$(2.8\pm 0.2) \left({\mid V_{cb}\mid \over 0.05}\right)^2$ &$3.1
\left({\mid V_{cb}\mid \over 0.05}\right)^2$\\
\& $m_c=1.27\, GeV$ & & & \\ \hline
\end{tabular} \end{center} \caption[]{{\small\sf This table lists three
of the five SM predictions made by SM case (a)  ($M_I=10^{10.94}\,
GeV$) and SM case (b) ($M_I=10^{14}\, GeV$) and those same three
parameters as predicted by the DHR ansatze with $M_S=180\, GeV$ and
$\alpha_3(M_Z)=0.121$. $M_I$ is the scale at which the intermediate
gauge symmetry $2_L\, 2_R\, 4_C$ breaks to the SM.}\label{SMpredict}}
\end{table}

\begin{table} \begin{center}
\begin{tabular}{|c||c|c|c|c|c|c|c|c|c|c|c|} \hline $U(1)$ & ${\bf
16_1}$ &
${\bf 16_2}$ & ${\bf 16_3}$ & ${\bf 10_1}$ & ${\bf 10_2}$ & ${\bf
10_3}$ &
${\bf \overline{126_1}}$ & ${\bf
\overline{126_2}}$ & ${\bf \overline{126_3}}$ & {\bf 45} & {\bf 210}\\
\hline \hline X & 1 &1 &1&-2&-2&-2&-2&-2&-2&0&0\\ \hline
Y&2&1&0&-3&-1&0&4&-2&0&$-{1\over 2}$&$-{1\over 2}$\\ \hline
Z&1&0&0&-1&0&0&-2&0&0&-1&0\\ \hline \end{tabular} \end{center}
\caption[]{{\small\sf Here we show how the three fermion SO(10) gauge
group spinor fields, three 126-dimensional representation Higgs fields,
the 45-dimensional Higgs field, and the 210-dimensional Higgs field of
our example model of Section 6 transform under the model's softly
broken three U(1) symmetries.}\label{inducedvevs}} \end{table}

\end{document}